\documentclass[journal]{IEEEtran}

\usepackage{xcolor,soul,framed} %,caption

\colorlet{shadecolor}{yellow}
\usepackage[pdftex]{graphicx}
\graphicspath{{../pdf/}{../jpeg/}}
\DeclareGraphicsExtensions{.pdf,.jpeg,.png}

\usepackage[cmex10]{amsmath}
\usepackage{array}
\usepackage{mdwmath}
\usepackage{mdwtab}
\usepackage{eqparbox}
\usepackage{url}

\begin{document}
\bstctlcite{IEEEexample:BSTcontrol}
    \title{Comment on ``An Efficient Privacy-Preserving Ranked Multi-Keyword Retrieval for Multiple Data Owners in Outsourced Cloud"}
  \author{Uma Sankararao Varri% <-this % stops a space

  \thanks{SRM University-AP, Amaravathi, Andhra Pradesh, India (umasankararao.v@srmap.edu.in)}}

% The paper headers

% ====================================================================
\maketitle

% === ABSTRACT ====================================================================
% =================================================================================
\begin{abstract}
%\boldmath
Protecting the privacy of keywords in the field of search over outsourced cloud data is a challenging task. In IEEE Transactions on Services Computing (Vol. 17 No. 2, March/April 2024), Li et al. proposed PRMKR: efficient privacy-preserving ranked multi-keyword retrieval scheme, which was claimed to resist keyword guessing attack. However, we show that the scheme fails to resist keyword guessing attack, index privacy, and trapdoor privacy. Further, we propose a solution to address the above said issues by correcting the errors in the important equations of the scheme. 
\end{abstract}

% === KEYWORDS ====================================================================
% =================================================================================

% For peer review papers, you can put extra information on the cover
% page as needed:
% \ifCLASSOPTIONpeerreview
% \begin{center} \bfseries EDICS Category: 3-BBND \end{center}
% \fi
%
% For peerreview papers, this IEEEtran command inserts a page break and
% creates the second title. It will be ignored for other modes.
\IEEEpeerreviewmaketitle

% === I. INTRODUCTION =============================================================
% =================================================================================
\section{Introduction}

\IEEEPARstart{T}{o} offer search over encrypted cloud data, searchable encryption (SE) has gained much attention from researchers all over the globe. In SE, data owners outsources encrypted documents along with encrypted index, where index contains the keywords which are extracted from the document collection and send to the cloud. Whenever user need to search a particular document, the user has to submit the trapdoor to the cloud which contains a keyword in encrypted form. Later, cloud performs a search between encrypted index and trapdoor, if any match found, it returns the documents to the user. The main goal of SE is to provide privacy-preservation of outsourced documents and keyword. In this setting, there are several schemes proposed from last two decades by enhancing the security and efficiency. These schemes are based on single keyword search and multi-keyword search (MKS). Single keyword search schemes contains only one keyword in the user trapdoor. However, with only one keyword, the accuracy in retrieving the documents will not be as expected. In multi-keyword search  schemes, user trapdoor contains more than one keyword. MKS is greatly exposed by the researchers because of its advantages. Recently, Li et al. \cite{Li2024} proposed a multi-keyword search scheme for multiple data owners to achieve privacy preserving ranked keyword search. Their main innovation is that the proposed scheme \cite{Li2024} is resistant to keyword guessing attacks and achieves privacy preservation of outsourced documents. Unfortunately, after closely observing the scheme \cite{Li2024}, we demonstrate that the scheme is easily vulnerable to keyword guessing attack of both indexed and trapdoor keywords. Further, the malicious cloud server can also view the plaintext of the documents. The detailed review of Li et. al.'s \cite{Li2024} scheme is shown in Section \ref{review}.
\section{Review of Li et. al.'s PRMKR Scheme}\label{review}
In this section, we review the important parts of algorithms of Li et. al.'s scheme \cite{Li2024}, the readers can refer \cite{Li2024} for complete understanding of the scheme. \\
$Setup(1^{\lambda}):$ Considering $\lambda$ as the security parameter, $g$ as the generator, $p$ as a large prime number, $e$ as a bilinear mapping function, $\{ G, G_{T} \}$ as cyclic groups, and $H:\{0,1\}^{*} \in Z_{p}$ as a hash function, the scheme produces the public parameters as
$Param = \{ G, G_{T}, e, p, g, H, k \}$, where $k$ is the document's secret key. \\
$KeyGen_{DO_{i}}(Params):$ By inputting public parameters, this algorithm, for each data owner i, selects a secret key $SK_{DO_{i}}$ $x_{i,t} \in Z_{p}$, $t:[1 - u]$, where $u$ is the number of keywords. Then computers a public key $PK_{DO_{i}}$ as $g^{x_{i,t}}$. \\
$KeyGen_{U_{j}}(Params):$ By inputting public parameters, this algorithm, for each data user j, selects a secret key $SK_{U_{j}}$ $y_{i,l} \in Z_{p}$, $l:[1 - q]$, where $q$ is the number of queried keywords. Then computers a public key $PK_{U_{j}}$ as $g^{y_{j,l}}$. \\
$IndexGen(Params, W, SK_{DO_{i}}, PK_{U_{j}}):$ By inputting the public parameters, keyword set $W$, secret key of data owner $SK_{DO_{i}}$, and public key of users $PK_{U_{j}}$, selects a random value $r_{i} \in Z_{p}$ for each $DO_{i}$. Then, the index is constructed as follows:
\[
I = \{ C_{i,1} = g^{H(w_{i,h}).x_{i,h}}, C_{i,2} = g^{(y_{j,h}).r_{i}}\}_{h = 1,2, \dots , u}
\]
$TrapdoorGen(Params, W^{\prime}, SK_{U_{j}}, PK_{DO_{i}}, r_{i}):$ By inputting the public parameters, query keyword set $W^{\prime}$, secret key of data user $SK_{U_{j}}$, public key of data owner $PK_{DO_{i}}$, and a random number $r_{i}$, the trapdoor is constructed as follows:
\[
T = \{ T_{j,1} = g^{H(w_{h^{\prime}}).y_{j,h^{\prime}}}, T_{j,2} = g^{(x_{i,h^{\prime}}).r_{i}}\}_{h^{\prime} = 1,2, \dots , q}
\]
$Search(I,T):$ By inputting index I and trapdoor T, the plug-in server computes the following Equation \ref{eq1}. If Equation \ref{eq1} holds, it returns the document IDs to user.
\begin{equation}\label{eq1}
    \frac{e(C_{i,1}, C_{i,2})}{e(T_{j,1}, T_{j,2})}
\end{equation}
\section{Analysis of Li et. al.'s PRMKR Scheme}
Though the scheme in \cite{Li2024} claimed to resist the keyword guessing attacks of indexed keywords and trapdoor keywords, and privacy-preservation of documents, we demonstrate how the scheme in \cite{Li2024} fails to hold the claims.  We show that the keyword information from both index and trapdoor can be extracted by launching offline keyword guessing attack. More specifically, any adversary or a plug-in server (since scheme \cite{Li2024} considered plug-in server as semi honest) can get the encrypted indexed keyword or encrypted trapdoor keyword information by launching offline keyword guessing attack (Attack I), and (Attack II), respectively. Furthermore, the correctness of search process (equation \ref{eq1}) is questionable because of the random number in the trapdoor. We demonstrate this in (Attack III). Finally, the malicious cloud can learn the plaintext form of the documents, as we shown in (Attack IV). The details of these four situations are as follows:  \\
\textit{Attack I:} Offline keyword guessing of Index I: For the given index $I = \{ C_{i,1}, C_{i,2}\}$, where $C_{i,1} = g^{H(w_{i,h}).x_{i,h}}$ and $C_{i,2} = g^{(y_{j,h}).r_{i}}$, any adversary can randomly select a keyword $w^{*}$ and check whether $w^{*}$ is hidden in the index I or not by computing the following: $e(C_{i,1}, g) = e(g^{H(w^{*})}, PK_{DO_{i}})$. \\
\textit{Attack II:} Offline keyword guessing of trapdoor T: For the given trapdoor $T = \{ T_{j,1}, T_{j,2}\}$, where $T_{j,1} = g^{H(w_{h^{\prime}})._{j,h^{\prime}}}$ and $T_{j,2} = g^{(x_{i,h^{\prime}}).r_{i}}$, any adversary can randomly select a keyword $w^{* \prime}$ and check whether $w^{* \prime}$ is hidden in the trapdoor T or not by computing the following: $e(T_{j,1}, g) = e(g^{H(w^{* \prime})}, PK_{U_{j}})$. \\
\textit{Attack III:} This is not the attack by the adversary but about the correctness of an important equation of the scheme. The search equation shown in Equation \ref{eq1} holds true, if $w_{i,h}$ is equals to $w_{h^{\prime}}$ and random component $r_{i}$ in $T_{j,2} = g^{(x_{i,h^{\prime}}).r_{i}}$ is equals to the random component $r_{i}$ in the index $C_{i,2} = g^{(y_{j,h}).r_{i}}$. However, $r_{i}$ is the secret component related to the data owner $DO_{i}$. Let use consider, \\
CASE I: If $r_{i}$ is shared to each data user, then there is no much use of index component $C_{i,2} = g^{(y_{j,h}).r_{i}}$ and trapdoor component $T_{j,2} = g^{(x_{i,h^{\prime}}).r_{i}}$, since they have only public information like $PK_{U_{j}}$ and $PK_{DO_{i}}$. Also If any malicious user misuses $r_{i}$ then the Equation \ref{eq1} acts more similar to $e(g^{H(w_{i,h}).x_{i,h}} , g) = e(g^{H(w_{h^{\prime}})._{j,h^{\prime}}},g)$. \\
CASE II: If $r_{i}$ in index and trapdoor is different, then there is no way that the Equation \ref{eq1} holds even if the keyword in the index and the trapdoor is same. $e(g^{H(w_{i,h}).x_{i,h}} , g^{(y_{j,h}).r_{i}}) = e(g^{H(w_{h^{\prime}})._{j,h^{\prime}}}, g^{(x_{i,h^{\prime}}).r_{i}^{*}})$ = $e(,g)g^{H(w_{i,h}).x_{i,h}(y_{j,h}).r_{i}} = e(g,g)^{H(w_{h^{\prime}})._{j,h^{\prime}}(x_{i,h^{\prime}}).r_{i}^{*}}$. Even if $w_{i,h}$ = $w_{h^{\prime}}$, $r_{i}$ should be equal to $r_{i}^{*}$. \\
\textit{Attack IV:} Documents were encrypted in the scheme using a symmetric key $k$. If $k$ is known to public then the cloud can easily access the plaintext information of any document which is stored in the cloud. There is a statement in this paper saying that ``the key $k$ can be transferred to the users securely". Unfortunately, key $k$ is included in the public parameters list  $Param = \{ G, G_{T}, e, p, g, H, k \}$. There is no difficulty for a malicious cloud to get any document's plaintext with $k$ as a public parameter. Hence, the document privacy is compromised in scheme \cite{Li2024}.
\section{Discussion and Solutions}
The security vulnerabilities of Li. et. al.'s schemes are mainly for the following reasons: 

\textit{Reason:} In the Index and Trapdoor generation phases, the keyword inclusion component $\{C_{i,1}, T_{j,1}\}$ are having all publicly available components $(g, H, g^{x_{i,h}}, g^{y_{j,h^{\prime}}})$ except the keyword. In such case, any adversary can easily guess the keyword by using the bilinear pairing between index/trapdoor component and publicly available components (As shown in Attack I and II). \\
\textit{Solution:} Create a master secret key (MSK) so that it can be included in $C_{i,1}$. Also whenever the secret key (SK) is generated to the user, include MSK in the SK such that the user cannot further extract the MSK from the SK. Then, user can use SK to generate $T_{j,1}$.
Adversaries cannot able to pair between $C_{i,1} / T_{j,1}$ and publicly available components unless then know the MSK. In this way, the index and trapdoor keyword guessing can be resisted. 

\textit{Reason:} The secret component $r_{i}$ of $DO_{i}$ is not useful if it is directly shared to the user because user may misuse it (As shown in Attack III). \\
\textit{Solution:} $r_{i}$ can be included in the MSK and it can be included in the user's secret key without the knowledge of $r_{i}$. On the other hand, $r_{i}$ in index and $r_{i}$ in trapdoor can be different, if adding more components in index and trapdoor. Later, the search equation must be changed for the correctness. 

\textit{Reason:} The document secret key is included in the public parameters list. By using this key, curious cloud may get the plaintext of the documents. \\
\textit{Solution:} The document secret key $k$ must be secret. It should be only with the users who are eligible to get the plaintext of the outsourced documents. Removing it from the public parameters list is a solution. But also sending it in a secure channel is a challenging issue. Attribute-based encryption \cite{bethencourt2007ciphertext} can be integrated with searchable encryption for efficient key management.
\section{Conclusion}
We analyzed the security aspects of PRMKR \cite{Li2024} scheme and showed that the major claims of the scheme are vulnerable in terms of security. We demonstrated that scheme in \cite{Li2024} is vulnerable to keyword guessing attacks and document privacy. Then, we have shown some possible solutions to overcome the vulnerabilities. In future, we will build an attribute-based keyword search scheme by addressing all the attacks demonstrated here.

% if have a single appendix:
%\appendix[Proof of the Zonklar Equations]
% or
%\appendix  % for no appendix heading
% do not use \section anymore after \appendix, only \section*
% is possibly needed

% use appendices with more than one appendix
% then use \section to start each appendix
% you must declare a \section before using any
% \subsection or using \label (\appendices by itself
% starts a section numbered zero.)
%

% ============================================
%\appendices
%\section{Proof of the First Zonklar Equation}
%Appendix one text goes here %\cite{Roberg2010}.

% you can choose not to have a title for an appendix
% if you want by leaving the argument blank
%\section{}
%Appendix two text goes here.

% use section* for acknowledgement
%\section*{Acknowledgment}

%The authors would like to thank D. Root for the loan of the SWAP. The SWAP that can ONLY be usefull in Boulder...

% Can use something like this to put references on a page
% by themselves when using endfloat and the captionsoff option.
\ifCLASSOPTIONcaptionsoff
  \newpage
\fi

% trigger a \newpage just before the given reference
% number - used to balance the columns on the last page
% adjust value as needed - may need to be readjusted if
% the document is modified later
%\IEEEtriggeratref{8}
% The "triggered" command can be changed if desired:
%\IEEEtriggercmd{\enlargethispage{-5in}}

% ====== REFERENCE SECTION

%\begin{thebibliography}{1}

% IEEEabrv,

\bibliographystyle{IEEEtran}
\bibliography{IEEEabrv,Bibliography}

\vfill

% Can be used to pull up biographies so that the bottom of the last one
% is flush with the other column.
%\enlargethispage{-5in}

% that's all folks
\end{document}